\documentclass[11pt]{article}
\usepackage{epsfig}
\usepackage{supercite}
\def\la{\mathrel{\hbox{\rlap{\hbox{\lower4pt\hbox{$\sim$}}}\hbox{$<$}}}}
\def\ga{\mathrel{\hbox{\rlap{\hbox{\lower4pt\hbox{$\sim$}}}\hbox{$>$}}}}

\title{\bf Detection of weak gravitational lensing distortions of distant
 galaxies by cosmic dark matter at large scales}

\author{David M. Wittman$^*$, J. Anthony Tyson, David Kirkman,\\ 
Ian Dell'Antonio$^\dagger$, and Gary Bernstein$^\ddagger$}

\begin{document}

\maketitle

\begin{center}
$^*$ Bell Laboratories, Lucent Technologies, Murray Hill, NJ 07574\\
$^\dagger$ Kitt Peak National Observatory, NOAO, Tucson, Arizona 85726\\
$^\ddagger$ Astronomy Department, University of Michigan, Ann Arbor, MI 48109 
\end{center}


\raggedright
\setlength{\parskip}{16pt}

{\bf Most of the matter in the universe is not luminous and can be
observed directly only through its gravitational effect.  
An emerging technique called weak gravitational lensing uses
background galaxies to reveal the foreground dark matter distribution
on large scales.  Light from very distant galaxies travels to us
through many intervening overdensities which gravitationally distort
their apparent shapes.  The observed ellipticity pattern of these
distant galaxies thus encodes information about the large-scale
structure of the universe, but attempts to measure this effect have
been inconclusive due to systematic errors.  We report the first
detection of this ``cosmic shear'' using 145,000 background galaxies
to reveal the dark matter distribution on angular scales up to half a
degree in three separate lines of sight. The observed angular
dependence of this effect is consistent with that predicted by two
leading cosmological models, providing new and independent support for
these models.}

\clearpage


The large-scale distribution of dark matter depends both upon the
nature of the dark matter and the global cosmological parameters that
describe the universe.  Information on the large scale distribution of
matter is thus one of the primary goals of modern observational
astronomy.  To date, most of what we know about the large-scale
structure of the universe comes from the observed anisotropies in the
cosmic microwave background (CMB) and from the distribution of
galaxies.  The CMB provides the earliest sample of mass fluctuations,
from a time when the universe was 100,000 times
younger.\cite{wilkinson} Different cosmological models predict
different scenarios in the growth of mass structures over cosmic time,
so comparison of the CMB-derived mass spectrum with that seen at later
times will be a powerful test of cosmology.  The large-scale mass
distribution at late times has traditionally been characterized
through the large-scale galaxy distribution, on the assumption that
light traces mass.

The distribution of this dark mass can be investigated more directly
via its gravitational effects on the appearance of background
galaxies.  Any foreground mass bends light rays from a distant source,
moving the apparent position of the source to a new position on the
sky and stretching its image tangentially, by an amount proportional
to the foreground mass.  This weak lensing effect has already been
used to study the mass distribution within clusters of galaxies, where
the large mass associated with the clusters makes the gravitationally
induced ellipticity of the background galaxies easily
detectable.\cite{tvw90,fahlman94,squires96,clowe98,hoekstra98,mellier99}
In principle, weak lensing can also tell us about large-scale
structure through the cumulative effect of many intervening
overdensities.  A deep image of a patch of the sky looks out through
the three-dimensional forest of galaxies seen in projection: any two
galaxies are most likely not physical neighbors and, absent lensing,
their projected shapes or ellipticities are statistically
uncorrelated.  In the presence of foreground mass overdensities, the
light rays from galaxies narrowly separated on the sky travel similar
paths past intervening mass concentrations and thus undergo similar
image distortions. The resulting correlation of distant galaxy
ellipticities is highest at small angular separation and drops for
widely separated galaxies whose light bundles travel through
completely different structures (Fig. 1).  Different cosmological
models predict different behavior for correlations of galaxy
ellipticites versus angle on the sky.

Theoretical expectations for this ``cosmic shear'' on 10-30
arcminute angular scales range from a
few percent for standard cold dark matter to less than one percent for
an open universe which would expand 
forever.\cite{gunn67,dyer74,jordi91,bland91,kaiser92,villumsen96,jain97,kaiser98a} 
The typical background galaxy has an intrinsic ellipticity of roughly
30\%, so that many thousands of source galaxies must be averaged
together to detect the small change induced by cosmic shear.  In
addition, a large area of sky must be covered, because mass structures
should span a few arcminutes to a degree at a typical mid-path
distance of redshift 0.4 (about three billion light-years).  Earlier
attempts to measure cosmic shear were 
inconclusive,\cite{kristian67,vtj83,mouldetal94,schneideretal98} 
with the main difficulty being
control of systematic errors in galaxy shapes arising from the optical
system or the process of observation.  The earliest attempts with
photographic plates, while covering a large field, suffered from
plate-to-plate systematics as well as nonlinearity and lack of
sensitivity.  The sensitivity, linearity, and reproducibility problems
were solved with the advent of charge-coupled devices (CCDs), but the
small field size covered by early CCDs was a problem.  Mosaics of
large CCDs now approach the desired one-degree field size and are
stimulating much activity in weak gravitational lensing.

We have imaged large areas of sky in several directions using a mosaic
of CCDs on a large telescope, covering hundreds of thousands of
distant galaxies at multiple wavelengths.  We describe the steps taken
to minimize systematic errors and to select 145,000 of the most
reliable distant galaxy measurements.  We find significant ellipticity
correlations on angular scales of $0.04^\circ-0.5^\circ$.  This is the
first direct probe of the aggregate mass distribution in the universe
on the several billion light-year scale, and the results are
consistent with two leading cosmological models.

\section*{Wide-field imaging with control of systematic shape errors}

We observed three ``blank'' ({\it i.e.}, not containing any known mass
concentrations) fields at 23$^h$48$^m$, +00$^\circ$57$^\prime$ J2000;
04$^h$29$^m$, -36$^\circ$18$^\prime$; and 11$^h$38$^m$,
-12$^\circ$33$^\prime$ over a period of several years, using the Big
Throughput Camera,\cite{wittmanetal98} an array of four large,
blue-sensitive CCDs at the Cerro Tololo Inter-American Observatory's
4-m Blanco telescope. Constructed originally for weak lens
observations, this camera covers a 35 arcminute field of view with
0.43 arcsecond pixels.  We took multiple 500 second exposures shifted
by 5-7 arcminutes and combined them to cover a 43 arcminute field.
Before combining, we took several steps to reduce systematic errors
arising from the optical system.  First, we registered all the images
onto a common linear coordinate system, free of the known radial
distortion of the telescope optics.  We then used the shapes of stars,
which are foreground point sources free of the gravitational lensing
effect, to correct any additional anisotropies in the point-spread
function (the response of the optical system to point sources), such
as those due to astigmatism and guiding errors. As described below,
our observations covered multiple wavelengths; this enables filtering
for certain types of stars and for distant galaxies.

The shape of a star or galaxy can be described by its second central moments,
$I_{xx} \equiv \Sigma Iwx^2$, $I_{yy} \equiv \Sigma Iwy^2$, and $I_{xy}
\equiv \Sigma Iwxy$, where $I(x,y)$ is the intensity distribution above
the night sky level, $w(x,y)$ is a weight function, the sum is over a
contiguous set of pixels defined as belonging to the galaxy, and the
coordinate system has been translated so that the first moments
vanish.  The second moments can be combined to form a size,
$I_{xx}+I_{yy}$, and two components of a pseudo-vector ellipticity,
$e_1 \equiv (I_{xx}-I_{yy}) / (I_{xx}+I_{yy})$ and $e_2 \equiv
2I_{xy}/(I_{xx}+I_{yy})$, which vary in the range $[-1,1]$
(ellipticity in its colloquial sense is the amplitude of this
pseudo-vector, $\epsilon \equiv \sqrt{e_1^2 + e_2^2}$ with its range
$[0,1]$).  Traditional intensity-weighted moments are calculated with
$w=1$, but this produces ellipticity measurements with noise
properties that are far from optimal or even divergent.  In cases of
white noise the formal optimal weight for an elliptical source is a
noise-free image of that elliptical source.\cite{castleman79}  In the
absence of such an image, weak lensing measurements are generally made
with circular Gaussian weights.  We use an elliptical Gaussian as the
weight function, which places more weight on the high-signal-to-noise
inner parts of the galaxy image, and is nearly optimal for most
point-spread functions and for typical exponential galaxy profiles.

The moments of the Gaussian weight ellipse are iterated (from initial
values provided by unweighted moments) to match the size and shape of
the object, in order to obtain the highest possible signal-to-noise
and to insure that the measured ellipticity is not biased toward the
shape of the weight function.  This ``adaptively weighted moments''
scheme has been extensively tested on simulated and real data, and has
been shown to be unbiased.  On simulated data, the algorithm recovers
a somewhat higher fraction of the artificially induced shear than does
simple intensity weighting.  However, our final results do not depend
on this particular weighting scheme.  Its real benefit lies in
rejection of peculiar objects, the vast majority of which are
overlapping galaxies seen in projection.  If the final centroid of a
galaxy differs significantly from the starting centroid, that galaxy
is rejected.  If the centroid does not shift, the galaxy is accepted
(and probably suffers little contamination by its neighbor).  Object
candidates are also rejected if the centroid or the ellipticity fails
to converge; if they are too near the edge of the image; if the size
grows too large; or if the moments are negative.  About one-third of
candidates found by the detection software (which provides the initial
unweighted moments, and can ``detect'' occasional noise peaks) are
rejected.  For candidates which survive, the measurement error in the
final ellipticity is accurately estimated by propagating the Poisson
photon noise through the moment equations.

We used foregound stars at many positions in the field of view to
measure and correct for systematic ellipticity error.  Stars
are distinguished from galaxies by their clear separation at the
bright end of a size--flux density diagram.  We identified roughly 100
such stars on each exposure of each CCD and made a least-squares fit
(with 3$\sigma$ clip) of a second-order polynomial to the spatial
variation of their ellipticity components $e_1^*(x,y)$ and
$e_2^*(x,y)$, which would be zero at all points in an ideal
observation free of point-spread function anisotropy.  Fischer and
Tyson\cite{ft97} have shown that nonzero $e_1^*$ and $e_2^*$ can be
cancelled by convolution with a small (three pixel square)
flux-conserving kernel with ellipticity components equal and opposite
to those of the stars. Simulations as well as weak lensing data on
clusters of galaxies show that faint galaxy induced systematics are
also removed in this process of circularising stars.  We convolved
each image with its resulting position-dependent circularising kernel,
after which the stellar ellipticities show little variation as a
function of position.  We then combined the images by
averaging with a 3$\sigma$ clip, and repeated the point-spread
function rounding on the combined image (using roughly 1000 stars and
a fourth-order polynomial in this case).  Fig. 2 depicts the
evolution of one of our worst raw images through this process.

\section*{Catalogues of distant galaxies}

We repeated the observing and image processing for each field in three
wavelength bands centered on 450 nm, 650 nm, and 850
nm,\cite{gullixsonetal95} and for two of the fields we also took 550
nm images.  The mean exposure time at each wavelength was 3400 sec.
In each field, we used standard software\,\cite{sextractor} to
identify object positions and fluxes on the 650 nm image (which is the
deepest image in each field), yielding roughly 150,000 objects per
field. We have confirmed the robustness of the weighted intensity
moments in our detected object catalogues by using different
(FOCAS\,\cite{focas} with adaptive circular kernel\,\cite{tyson94})
detection and evaluation software.  At each object's position, we
evaluated the weighted moments at each wavelength, retaining only the
measurements which the iterative weighted moment algorithm did not
flag as suspect.  Measurements with small sizes ($I_{xx} < 1.0$ or
$I_{yy}<1.0$) were also excluded as suspect.  The result is a list of
multiple independent ellipticity measurements (and corresponding
estimated measurement errors $\sigma_i$) for each object.

We then computed the best estimate of each galaxy's ellipticity by
averaging the remaining measurements, weighted inversely by their
estimated errors.  If either of the ellipticity components at any
wavelength deviated from this mean by more than $3\sigma_i$, that
wavelength was eliminated and the process repeated.  This step thus
eliminates individual galaxy ellipticity measurements at wavelengths
at which objects were noisy or blended, and it also reduces the
systematic errors because the images at different wavelengths do not
share the same residual point-spread function anisotropy.  Finally we
rejected objects with $\epsilon>0.6$ as likely to be blends of more
than one object, and applied flux density criteria ($1.8\,\mu Jy >
F_\nu > 0.11\,\mu Jy$ through the 650 nm filter, 23--26 R magnitude)
to select objects likely to be distant galaxies. We use these same
selection criteria in calibrating the typical redshift of the background
galaxies (below).  The final catalogues contain about 45,000 galaxies
in each field.  A visual inspection of the final catalogues indicates
that they are free of spurious objects such as bits of scattered light
around bright stars.

These observed ellipticities must be corrected for the overall
broadening effect of the point-spread function, which makes elliptical
galaxies appear more circular even if the point-spread function itself
is perfectly isotropic.  To calibrate this effect, we took a deep
image with a very small point spread (the Hubble Deep Field South) and
convolved it to the resolution of our final images, which is
1.07--1.25 arcsec as measured by the full width at half-maximum at 650
nm.  (The resolution on individual exposures, or ``seeing'' was
better, but the stellar size is larger in the final image with
systematic shape errors removed from the point-spread function.)
While some isolated galaxies became broader and less elliptical as
predicted, most merged with their neighbors, producing many more
elliptical objects than predicted and preventing the construction of a
clear relationship between observed and true ellipticity for
individual galaxies.  

Instead we calibrated the fraction of cosmic shear recovered, as a
function of resolution, from the ensemble of galaxies matching our
selection criteria.  We induced a known shear into the Hubble Deep
Field South, convolved to the desired resolution, applied the same
galaxy measurement and selection routines (at 650 nm only), and
measured the mean ellipticity of the resulting sample.  We averaged
over repeated shears in several different directions to assess the measurement
errors.  The ratio of induced to recovered ellipticity was $4.5 \pm
0.5$, with no clear trend as a function of resolution.  The lack of
such a trend would be quite surprising for isolated galaxies, but the
coalescence of galaxy images appears to be the dominant effect.  In
the fairly small range of 1.07--1.25 arcsec resolution, this effect
does not change the recovery factor by more than the measurement error
of 0.5, so we adopt 4.5 as an overall ellipticity recovery factor.

\section*{Ellipticity correlations of distant galaxies}

Miralda-Escud\'{e}\,\cite{jordi91} has defined two physically
revealing ellipticity correlation functions.  In this approach, the
ellipticity components of a galaxy $i$ are calculated not with respect
to the arbitrary $x$ and $y$ axes of the image, but with respect to
the line joining it to another galaxy $j$ (Fig. 3).  Averaging over
all galaxies $i$ and $j$ separated by angle $\theta$ on the sky, the
correlations $\xi_1(\theta) \equiv \langle e_{1i} e_{1j}\rangle$ and
$\xi_2(\theta) \equiv \langle e_{2i} e_{2j}\rangle$ have a unique
signature in the presence of gravitational lensing, explained in
detail in Fig. 3.  We recently reported the detection of a cosmic shear
signal in the quadrature sum of these correlation functions.\cite{wt99}

Fig. 4 shows the ellipticity correlations for each of the three
fields in the angular separation range 2-36 arcmin (top panels).  The
plotted errors indicate 68\% confidence intervals determined from 200
bootstrap-resampled realizations of the final galaxy catalogue in each
field.  Note that the measurements in different angular separation
bins are not statistically independent, but $\xi_1$ and $\xi_2$ are
independent from each other, as are the three fields.  At $\theta =
6.1$ arcmin, the confidence that $\xi_1 > 0$ is 97\%, 99.5\%, and
99.5\% for the three fields in the order shown in Fig. 4.
Similarly, the confidence that $\xi_2 > 0$ at the same angular scale
is 87.5\%, $>99.5\%$, and 97\% respectively.  Some ``cosmic
variance'', or real systematic differences among fields of this size,
is expected,\cite{kruse99} but the statistical errors are too large to
examine this effect.  We plot the average over the three fields in the
lower panels of Fig. 4, with $1\sigma$ errors in the mean derived
from the variance among the fields (black points and errors).  The
signature of gravitational lensing by large-scale structure is
evident: $\xi_1$ declines as the angular scale increases, but is
positive at all scales, while $\xi_2$ matches $\xi_1$ at small scales
but drops below zero at large scales.  This result is robust: Similar,
but lower signal-to-noise, profiles are obtained if we use unweighted
moments or moments from the 650 nm images only.

We performed several tests for systematic errors.  The effects of
residual point-spread function anisotropy are demonstrated by plotting
the correlation functions of the stars (blue in Fig. 4). These are far
closer to zero than are the galaxy correlations.  Only $\xi_1$ in the
innermost bin has an apparently significant stellar correlation.  To
test the effect that this might have on the galaxy correlations, we
computed the star-galaxy correlations $\langle
e_{1,star}e_{1,gal}\rangle$ and $\langle e_{2,star}e_{2,gal}\rangle$
(green in Fig. 4).  The star-galaxy correlations are extremely close to
zero in this bin.  There are also tests involving the galaxy sample
alone. The cross correlation $\xi_3 \equiv \frac{1}{2}(\langle
e_{1i}e_{2j}\rangle + \langle e_{2i}e_{1j}\rangle)$ should vanish in
the absence of systematic errors (red in Fig. 4).  The result is
reassuringly close to zero.  The plotted errors for $\xi_3$ can also
be taken as an indicator of the statistical error associated with the
number and distribution of galaxies included in the catalogues (but
reduced due to the averaging of two functions in $\xi_3$).  This
estimate of statistical error agrees roughly with that shown for
$\xi_1$ and $\xi_2$.  Finally, the weak lensing signature disappears
if we randomise the galaxy positions.  

Apart from these null tests, there are also affirmative tests.  One
test is to take similar data centered on a cluster of galaxies of
known mass. A 650-nm image of massive cluster at redshift 0.45, taken
with the same camera and processed in the same way, exhibits
correlation functions ($\xi_1$ and $\xi_2$) of the expected angular
dependence and of larger amplitude than in any of the three blank
fields, despite likely contamination of the galaxy sample by cluster
members.  $\xi_3$ also vanishes in this field.  Another test involves
inverting the background galaxy ellipticity distribution to yield a
map of projected mass in each ``blank'' field.  We find occasional
mass concentrations which can often be identified with likely
foreground clusters, but no linear features or pileups at the edges of
the image which might indicate problems in the background galaxy
catalogues.  Furthermore, when a mass map is made using only those
galaxies likely to be behind a serendipitous cluster (based on colour
information), the lensing signal from that cluster increases markedly.
This corroborates the idea that the correlation functions are
accumulating over many sources and many overdensities spread
throughout the line of sight.  All these tests indicate that we have
indeed measured cosmic shear in our ``blank'' fields and that
contamination from surviving systematic error is low.  We now turn to
comparisons with theoretical predictions of this effect.

\section*{Comparison with theoretical predictions}

%
%
Ellipticity correlations increase strongly with background galaxy
redshift, so we must first constrain the source redshift distribution
$N(z)$.  Very little is known about the redshift distribution of
galaxies as faint as those used here, so we assume a simple model
$N(z) \propto z^2~exp(-z/z_0)$, and adjust $z_0$ to match weak
gravitational lensing observations of a high-redshift galaxy cluster
of known velocity dispersion (MS1054 at $z=0.83$).\cite{tranetal99} We
observed this cluster with the same camera and telescope and reduced
the data in the same way as for the blank fields, and compared the
faint galaxy ellipticities (tangential to the cluster center) to that
expected for a range of $z_0$.  We found that $z_0 = 0.5$ was the best
match.

This model $N(z)$ was used as input to a cold dark matter simulation
code by W. Hu and J. Miralda-Escud\'{e} (see ref
\cite{hu99}), which computes the shear power spectrum 
and correlation function for any given cosmology, using the
prescription of Hamilton et al.\cite{hamilton91} and Peacock \&
Dodds\cite{peacock96} to calculate the mass power spectrum in the
non-linear regime when the growth of gravitationally collapsed dark
matter structures modified the mass spectrum. Results were obtained
for three cosmological models and are plotted along with our
seeing-corrected measurements on a logarithmic scale in Figure 5.  Two
current models were normalised to the microwave background
fluctuations ({\it COBE}) at large angle and to local galaxy cluster
abundance (assuming mass traces light) at small angle: an open
universe with $\Omega_{\rm matter}=0.45$ (orange in Figure 5), and a
flat universe dominated by a cosmological constant $\Lambda = 0.67$
(green, solid line). The agreement between the data and the two viable
cosmological models is impressive for a first measurement.  For
comparison purposes we also show the old standard cold dark matter
flat cosmology (blue), which is only {\it COBE} normalised.  (A full
listing of the parameters used in these models is shown in Table 1.)
To illustrate the effect of varying $N(z)$, we also plot the
$\Lambda$-dominated cosmology with $z_0=0.3$ (green, dotted line).
Since our model $N(z)$ peaks at $z = 2 z_0$, this lowers the typical
redshift from 1.0 to 0.6.  Decreasing $z_0$ decreases the amplitude of
the correlations, but has little effect on their shapes.  The
uncertainty in $N(z)$ implies a factor of several uncertainty in the
amplitude of the correlation, and is by far the dominant calibration
error.

Despite this uncertainty, {\it COBE}-normalised standard cold dark
matter is ruled out by the measured values of $\xi_1$.  While this is
not surprising, it is the first cosmological constraint from
wide-field weak lensing, and it agrees with several other methods
which disfavor this model.\cite{ostriker95,bahcall99} The other two
models are consistent with the data at the $3\sigma$ level.  The
indication of a low $\Omega_{\rm matter}$ universe here is in
agreement with a remarkable array of independent methods, including
type Ia supernovae, cosmic microwave background anisotropies, cluster
baryon fraction together with cluster mass (lensing) and primeval
deuterium, and the age of the oldest stars coupled with the Hubble
constant\cite{tt99}.  However, the shape of $\xi_2$ is not a good fit
to either of these two model cosmologies, which are based on a single
power-law mass spectrum. If confirmed by further data, this would
suggest the need for a more complicated mass spectrum.

This technique can further distinguish between open and
$\Lambda$-dominated universes if extended to the somewhat larger
angular scales where those cosmologies predict $\xi_2$ will cross zero
as shown in Figure 5.  A survey of many $2^\circ \times 2^\circ$ fields now
underway will rule out one or more of these cosmologies at the $\sim 8\sigma$
level at 10 arcmin angles ($3\sigma$ level for a differential
measure of the slope of the power spectrum).  
Separating the background galaxies into discrete redshift bins based
on multi-colour photometry will enable measurement of the
ellipticity correlation (or equivalently the dark matter power
spectrum) as a function of cosmic time;
wide-field weak lensing surveys deep enough to identify
galaxies at $z \sim 2$ and measure their shapes will constrain several
cosmological parameters.\cite{hu99} Ultimately, the combination of all 
the power spectrum probes (lensing, cosmic microwave background, galaxy
distributions, and peculiar velocities) will tightly constrain
theories of the origins of fluctuations in the early universe and
their growth into galaxies and large-scale structure.

%
%
\subsection*{Acknowledgements}
We gratefully acknowledge help from Wayne Hu and Jordi
Miralda-Escud\'{e} on theoretical predictions of several cosmological
models.  We thank Steven Gentile for his artwork, and the staff of
CTIO for their help with the BTC project and for their upgrading and
maintenance of the delivered image quality of the Blanco telescope.
Cerro Tololo Inter-American Observatory is a division of National
Optical Astronomy Observatory (NOAO), which is operated by the
Association of Universities for Research in Astronomy, Inc., under
Cooperative Agreement with the National Science Foundation.  BTC
construction was partially funded by the NSF.

\bibliographystyle{unsrt}
\bibliography{ocf}

\begin{thebibliography}{10}

\bibitem{wilkinson}
L.~Page and D.~T. Wilkinson.
\newblock The cosmic microwave background.
\newblock {\em Rev. Mod. Phys.}, 71:173--179, 1999.

\bibitem{tvw90}
J.~A. Tyson, F.~Valdes, and R.~Wenk.
\newblock Detection of systematic gravitational lens galaxy image alignments:
  mapping dark matter in galaxy clusters.
\newblock {\em Astrophys. J.}, 349:L1--L4, 1990.

\bibitem{fahlman94}
G.~Fahlman, N.~Kaiser, G.~Squires, and D.~Woods.
\newblock Dark matter in ms1224 from distortion of background galaxies.
\newblock {\em Astrophys. J.}, 437:56--62, 1994.

\bibitem{squires96}
G.~Squires, N.~Kaiser, G.~Fahlman, A.~Babul, and D.~Woods.
\newblock A weak gravitational lensing analysis of abell 2390.
\newblock {\em Astrophys. J.}, 469:73--77, 1996.

\bibitem{clowe98}
D.~Clowe, G.~A. Luppino, N.~Kaiser, J.~P. Henry, and I.~M. Gioia.
\newblock Weak lensing by two $z \sim 0.8$ clusters of galaxies.
\newblock {\em Astrophys. J.}, 497:61--64, 1998.

\bibitem{hoekstra98}
H.~Hoekstra, M.~Franx, K.~Kuijken, and G.~Squires.
\newblock Weak lensing analysis of cl 1358+62 using hubble space telescope
  observations.
\newblock {\em Astrophys. J.}, 504:636--660, 1998.

\bibitem{mellier99}
Y.~Mellier.
\newblock Probing the universe with weak lensing.
\newblock {\em Annu. Rev. Astron. Astrophys.}, 37:127--189, 1999.

\bibitem{gunn67}
J.~E. Gunn.
\newblock A fundamental limitation on the accuracy of angular measurement in
  observational cosmology.
\newblock {\em Astrophys. J.}, 147:61--72, 1967.

\bibitem{dyer74}
C.~Dyer and R.~Roeder.
\newblock Observations in locally inhomogeneous cosmological models.
\newblock {\em Astrophys. J.}, 189:167--175, 1974.

\bibitem{jordi91}
J.~Miralda-Escud\'{e}.
\newblock The correlation function of galaxy ellipticities produced by
  gravitational lensing.
\newblock {\em Astrophys. J.}, 380:1--8, 1991.

\bibitem{bland91}
R.~Blandford, A.~Saust, T.~Brainerd, and J.~Villumsen.
\newblock The distortion of distant galaxy images by large scale structure.
\newblock {\em Mon. Not. Royal Astron. Soc.}, 251:600--627, 1991.

\bibitem{kaiser92}
N.~Kaiser.
\newblock Weak gravitational lensing of distant galaxies.
\newblock {\em Astrophys. J.}, 388:272--286, 1992.

\bibitem{villumsen96}
J.~Villumsen.
\newblock Weak lensing by large-scale structure in open, flat and closed
  universes.
\newblock {\em Mon. Not. Royal Astron. Soc.}, 281:369--383, 1996.

\bibitem{jain97}
B.~Jain and U.~Seljak.
\newblock Cosmological model predictions for weak lensing.
\newblock {\em Astrophys. J.}, 484:560--573, 1997.

\bibitem{kaiser98a}
N.~Kaiser.
\newblock Weak lensing and cosmology.
\newblock {\em Astrophys. J.}, 498:26--42, 1998.

\bibitem{kristian67}
J.~Kristian.
\newblock On the cosmological distortion effect.
\newblock {\em Astrophys. J.}, 147:864--867, 1967.

\bibitem{vtj83}
F.~Valdes, J.~A. Tyson, and J.~F. Jarvis.
\newblock Alignment of faint galaxy images: cosmological distortion and
  rotation.
\newblock {\em Astrophys. J.}, 271:431--441, 1983.

\bibitem{mouldetal94}
J.~Mould, R.~Blandford, J.~Villumsen, T.~Brainerd, I.~Smail, T.~Small, and
  W.~Kells.
\newblock A search for weak distortions of distant galaxy images by large-scale
  structure.
\newblock {\em Mon. Not. Royal Astron. Soc.}, 271:31--38, 1994.

\bibitem{schneideretal98}
P.~Schneider, L.~van Waerbeke, Y.~Mellier, B.~Jain, S.~Seitz, and B.~Fort.
\newblock Detection of shear due to weak lensing by large-scale structure.
\newblock {\em Astron. Astrophys.}, 333:767--778, 1998.

\bibitem{wittmanetal98}
D.~Wittman, J.~A. Tyson, G.~M. Bernstein, R.~W. Lee, I.~P. Dell'Antonio,
  P.~Fischer, D.~R. Smith, and M.~M. Blouke.
\newblock Big throughput camera: the first year.
\newblock {\em Proc. Soc. Photo-Optical Instr. Eng.}, 3355:626--634, 1998.

\bibitem{castleman79}
K.~R. Castleman.
\newblock Digital image processing.
\newblock {\em Prentice Hall}, page 214, 1979.

\bibitem{ft97}
P.~Fischer and J.~A. Tyson.
\newblock The mass distribution of the most luminous x-ray cluster
  rxj1347.5-1145 from gravitational lensing.
\newblock {\em Astron. J.}, 114:14--24, 1997.

\bibitem{gullixsonetal95}
C.~A. Gullixson, P.~C. Boeshaar, J.~A. Tyson, and P.~Seitzer.
\newblock The $b_jri$ photometric system.
\newblock {\em Astrophys. J. Supp.}, 99:281--293, 1995.

\bibitem{sextractor}
E.~Bertin and S.~Arnouts.
\newblock Sextractor: software for source extraction.
\newblock {\em Astron. Astrophys. Supp.}, 117:393--404, 1996.

\bibitem{focas}
F.~Valdes.
\newblock Resolution classifier.
\newblock {\em Soc. Photo-Optical Instr. Eng. (SPIE) Proceedings},
  331:465--472, 1982.

\bibitem{tyson94}
J.~A. Tyson.
\newblock Dark matter mapping by gravitational lens tomography.
\newblock {\em AIP Conf. Proc. {\it Dark Matter}, Eds: S. Holt and C. Bennett},
  (AIP Press 1995):287--296, 1995.

\bibitem{wt99}
D.~Wittman and J.~A. Tyson.
\newblock The shear correlation function out to 20 arcminutes.
\newblock {\em Gravitational Lensing: Recent Progress and Future Goals}, eds.
  T. G. Brainerd and C. S. Kochanek (ASP Conference Series):in press, 2000.

\bibitem{kruse99}
G.~Kruse and P.~Schneider.
\newblock The non-gaussian tail of cosmic shear statistics.
\newblock {\em astro-ph/9904192}, 1999.

\bibitem{tranetal99}
K.~H. Tran, D.~D. Kelson, P.~van Dokkum, M.~Franx, G.~D. Illingworth, and
  D.~Magee.
\newblock The velocity dispersion of ms1054-03: a massive galaxy cluster at
  high redshift.
\newblock {\em Astrophys. J.}, 522:39--45, 1999.

\bibitem{hu99}
W.~Hu.
\newblock Power spectrum tomography with weak lensing.
\newblock {\em Astrophys. J.}, 522:L21--L24, 1999.

\bibitem{hamilton91}
A.~J.~S. Hamilton, A.~Matthews, P.~Kumar, and E.~Lu.
\newblock Reconstructing the primordial spectrum of fluctuations of the
  universe from the observed nonlinear clustering of galaxies.
\newblock {\em Astrophys. J.}, 374:L1--L4, 1991.

\bibitem{peacock96}
J.~A. Peacock and S.~J. Dodds.
\newblock Non-linear evolution of cosmological power spectra.
\newblock {\em Mon. Not. Royal Astron. Soc.}, 280:L19--L26, 1996.

\bibitem{ostriker95}
J.~P. Ostriker and P.~J. Steinhardt.
\newblock The observational case for a low density universe with a cosmological
  constant.
\newblock {\em Nature}, 377:600--602, 1995.

\bibitem{bahcall99}
N.~A. Bahcall, J.~P. Ostriker, S.~Perlmutter, and P.~J. Steinhardt.
\newblock The cosmic triangle: revealing the state of the universe.
\newblock {\em Science}, 284:1481--1488, 1999.

\bibitem{tt99}
M.~S. Turner and J.~A. Tyson.
\newblock Cosmology at the millennium.
\newblock {\em Rev. Mod. Phys.}, 71:145--164, 1999.

\bibitem{frenk90}
C.~Frenk, S.~D.~M. White, G.~Efstathiou, and M.~Davis.
\newblock Galaxy clusters and the amplitude of primordial fluctuations.
\newblock {\em Astrophys. J.}, 351:10--21, 1990.

\end{thebibliography}

\clearpage
\pagestyle{empty}

\setlength{\textwidth}{8in}
\setlength{\oddsidemargin}{0in}
\setlength{\evensidemargin}{0in}
\vspace{10pt}
\begin{tabular*}{6.2in}[t]{l} \hline
{\bf  Table 1 Summary of cosmological models}\\ 
\end{tabular*}
\begin{tabular*}{6.2in}[t]{llclllll} \hline
Model (Fig 5 colour) & $\Omega_b$ & $\Omega_{\rm matter} - \Omega_b$ & $\Omega_\Lambda$ &  $H_0$  & n & $\sigma_8$ & normalization \\  \hline
Standard cold dark matter (blue)  & 0.05 & 0.95 & 0 &  50   &  1.0   &  1.17  &  {\it COBE} only \\
$\Lambda$-dominated, flat (green)  & 0.039 & 0.291 & 0.67 &  70  & 0.94  &  0.84  & {\it COBE}+clusters \\
Open universe (orange)  & 0.045 & 0.405 & 0 &  65  &  1.01 & 0.71  &  {\it
COBE}+clusters \\  \hline \\
\end{tabular*}
These cosmological
models were chosen in order to put our ellipticity correlation
measurements in context. The old standard cold dark matter model in
which the universe is nearly closed by cold dark matter, is also
disfavored in other observations. Its rms mass contrast is normalized
to that found 300,000 years after the Big Bang via the cosmic
microwave background radiation fluctuations observed with the 
COsmic Background Explorer satellite ($COBE$). The other two
models agree with a wide variety of observations, but only the
cosmological constant ($\Lambda$-dominated, flat)
cosmology also agrees with the recent evidence from
supernova studies for accelerated expansion.  $\Omega_b$ is the
fraction of critical density in to ordinary (baryonic) matter.
$\Omega_{\rm matter}$ is the fraction in all matter (mostly dark matter), and
$\Omega_\Lambda$ is the fraction in dark energy (the cosmological
constant).  $H_0$ is the Hubble constant in units of km s$^{-1}$
Mpc$^{-1}$.  The power spectrum $P(k)$ of mass density fluctuations 
is often plotted in
terms of inverse size: the wave number k is inversely proportional to length.
The parameter n is the slope of the primeval density power spectrum as a 
function of k: $P(k)\propto k^n$. For a scale-free power
spectrum of density fluctuations, n = 1. The parameter $\sigma_8$ is the current 
rms mass contrast in a random sphere of radius 8($100/H_0$) Mpc compared
with that for numbers of galaxies.\protect\cite{frenk90} The choice of n=1 and
$COBE$ normalization for standard cold dark matter results in too much
mass fluctuation on galaxy cluster scales.  By adjusting the slope
n and current rms mass contrast $\sigma_8$, models can be forced to
fit the rms mass contrast now on galaxy cluster scales as well as the
$COBE$ normalization.

\clearpage
\pagestyle{empty}
\setlength{\textwidth}{6.2in} 
\setlength{\oddsidemargin}{0.5in}
\setlength{\evensidemargin}{0.5in}

%
%
\begin{figure}

Please get 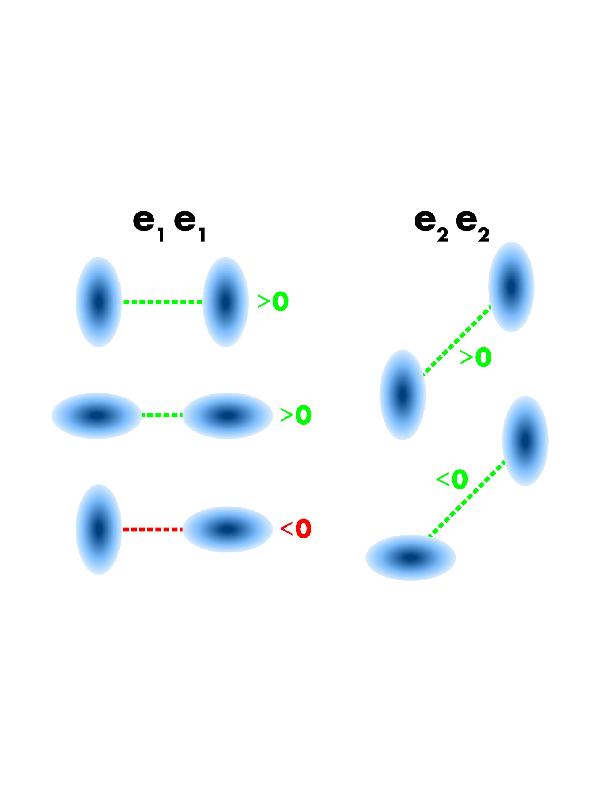 from astro-ph/0003014.

\caption{{\bf The distorted universe.}
Light rays from distant galaxies travel a tortuous path through
a universe filled with clustering dark mass. Every bend in the path
of a bundle of light from a distant galaxy stretches its apparent
image. The orientation of the resulting elliptical images of galaxies
contains information on the size and mass of the gravitational lenses
distributed over the light path. The figure shows a schematic view of 
weak gravitational lensing by large-scale mass structure: distant galaxy 
orientation is correlated on scales characteristic of the lensing dark
matter structures.  Light bundles from two distant
galaxies which are projected closely together on the sky follow
similar paths and undergo similar gravitational deflections by
intervening dark matter concentrations. Apparent orientations of
distant galaxies are thus correlated on angular scales of less than
a few degrees.  The larger the mass in the gravitational deflectors, the
larger the faint galaxy ellipticity correlations on a given angular scale.
These ellipticity correlations of distant galaxies reveal the
statistics of the large-scale dark mater distribution in the intervening
universe -- a key diagnostic of the underlying cosmology.
}
\end{figure}
\clearpage

%
%
\clearpage
\begin{figure}
\centerline{\psfig{file=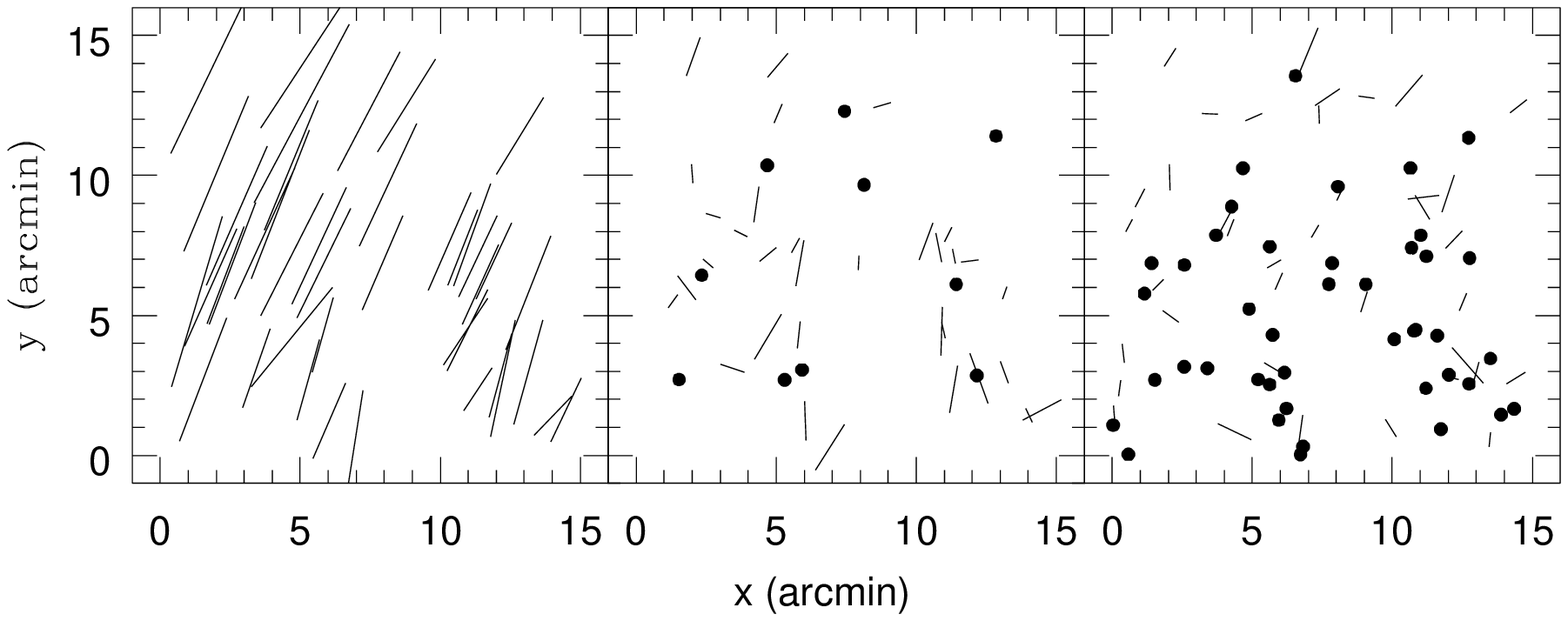,width=5in}}
\caption{{\bf Making stars round.} Foregound stars at many positions
in the field of view are used to correct for position-dependent
systematic ellipticity error.  Convolution with a position-dependent
kernel with ellipticity components equal and opposite to those of the
stars, reduces this systematic error everywhere in the field.  Here we
illustrate this technique with one particularly bad frame of raw data
from one of the four CCDs in our mosaic.  Each panel represents stars
at their positions in the field as a line encoding the ellipticity and
position angle, or as a point if the ellipticity is less than one
percent.  The left panel is the raw data; the stars in a more typical
frame have only half the ellipticity of those shown here, or about
5\%.  The middle panel shows the stellar shapes in the same single
image after convolution with the rounding kernel.  The stars are
vastly less out of round, but local correlations still exist.  The
right panel shows the final shapes of stars in the same region of sky,
after combining ten shifted exposures, convolving the combined image,
and measuring the shapes in more than one filter.  Many of the
local correlations in the middle panel have disappeared.  The density
of stars is greater due to better identification of stars in the
combined image.  The final field size is roughly ten times larger in
area than this patch, and contains about 1000 such stars in most of
our mid-latitude fields. This figure is for illustration purposes
only; Figure 4 contains a quantitative assessment of the final level of
systematic error.}
\end{figure}
\clearpage

%
%
\begin{figure}

Please get fig3.jpg from astro-ph/0003014.

\caption{{\bf Lens-induced galaxy orientation correlations.}  Pairs of
background galaxies, separated on the sky by some angle, can have
their relative orientations affected by weak lensing.  The ellipticity
components of any galaxy $i$ with respect to galaxy $j$ can be
visualised as $e_1 = \epsilon \cos(2\theta)$ and $e_2 = \epsilon
\sin(2\theta)$, where $\epsilon$ is the scalar ellipticity and $\theta$ is
the position angle with respect to a line joining the two galaxies.
Ellipticity correlation functions are computed from the products of the
ellipticity components of millions of such pairs, as a function of
angular separation between pairs.  A variety of relative orientations
are illustrated along with their contributions to the correlations.
Gravitational lensing leaves its signature on these correlations in
several ways.  First, the amplitude of the correlations scales with
the amount of foreground mass.  Second, the correlations are large at
small separations and drop to nearly zero at large separations in a
particular way.  The bottom (red) case on the left cannot be caused by
gravitational lensing, so that $\langle e_1e_1 \rangle$ (averaged over
many pairs) is always positive in the absence of systematic error.
But lensing can cause the $e_2$ product to have either sign, as shown,
and $\langle e_2e_2 \rangle$ should become negative at a separation
characteristic of the underlying cosmology.  Third, correlations
between $e_1$ and $e_2$ (not shown here) are not induced by
gravitational lensing, so that any putative measurement of a weak 
lensing effect should vanish in the cross correlation
$\langle e_1 e_2\rangle + \langle e_2 e_1\rangle$.
}
\end{figure}
\clearpage

%
%
\clearpage
\begin{figure}
\vspace*{-1in}
\centerline{\psfig{file=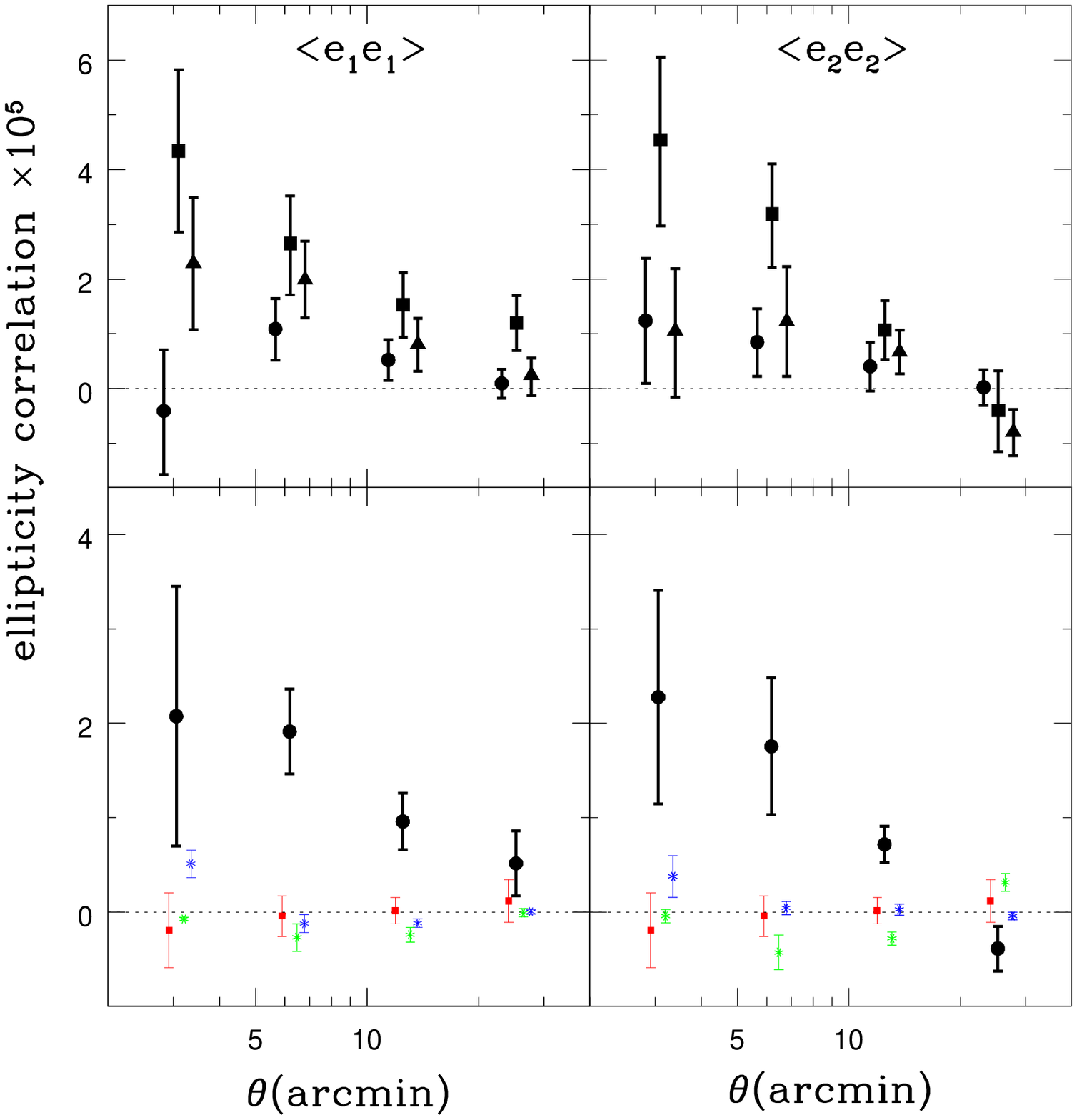,height=5in,width=5in}}
\caption{{\bf Detection of ellipticity correlations.}  The upper
panels show the measured ellipticity correlations as a function of
angle for three independent fields covering a total of 1.5 square
degrees ($\xi_1$ at left and $\xi_2$ at right).  Markers have been
slightly offset horizontally for clarity.  From left to right in each
bin are fields at 11$^h$38$^m$, -12$^\circ$33$^\prime$, 23$^h$48$^m$,
+00$^\circ$57$^\prime$, and 04$^h$29$^m$, -36$^\circ$18$^\prime$
(J2000).  In each field, roughly 45,000 faint galaxies passed all the
filters and significance tests, from an initial catalogue of about
150,000 objects.  Errors shown are 68\% confidence intervals
determined from 200 bootstrap-resamplings of the galaxy catalogues.
The lower panels show the mean of the ellipticity correlation
functions over the three fields (black), with 1$\sigma$ errors derived
from the variance between fields.  The behavior as a function of angle
matches that expected from weak gravitational lensing by large-scale
structure.  The lower panels also contain several null tests of
systematic error.  The cross-correlation of the galaxies $\xi_3$
should vanish in the absence of systematic error, and in fact is
everywhere consistent with zero (red).  The ellipticity correlations
of stars (blue) are everywhere consistent with zero except in the
innermost bin of $\xi_1$.  The effect of nonzero stellar correlations
on the galaxy correlations is illustrated by the star-galaxy
correlation (green), which is very close to zero in this bin.  An
additional test of systematics, a search for preferential alignment of
galaxies with the CCD axes, is also null.  Though galaxy ellipticity
correlations continue to rise at smaller angles, the smaller number of
galaxy pairs makes the measurement noisier, there are few closely-spaced
stars to assess systematic error, and the
theoretical interpretation on small scales is difficult.}
\end{figure}
\clearpage

%
%
\clearpage
\begin{figure}
\vspace*{-0.5in}
\centerline{\psfig{file=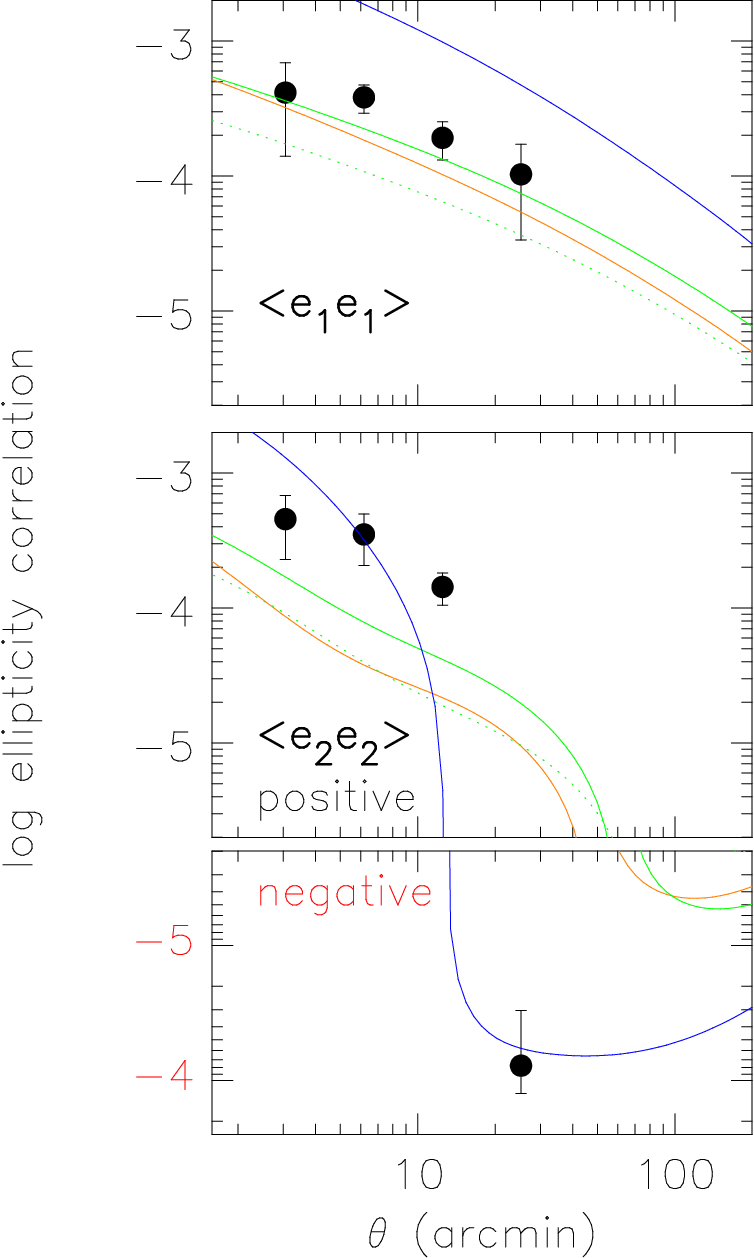,height=5in}}
\caption{{\bf Comparison of ellipticity correlations with
predictions.}  We plot our measurements with 1$\sigma$ errors on a
logarithmic scale along with theoretical predictions based on various
models for a cold dark matter universe.  The top theoretical curve is
for the old standard cold dark matter model (blue) and the center and
lower curves are for a universe with a cosmological constant
($\Lambda$CDM, solid green) and an open universe (orange),
respectively.  The dotted green curve shows the effect of decreasing
the mode of the background galaxy redshift distribution, $2z_0$, from
1.0 to 0.6 for one model ($\Lambda$CDM).  The errors shown are derived
from the variance among three fields; the statistical errors in each
field are larger than any cosmic variance present.  The data from
Figure 4 have been multiplied by a correction factor of 20 here, to
compensate for the ellipticity dilution factor of 4.5 described in the
text, which is squared in the correlation functions.  The measurements
are consistent with $\Lambda$CDM and an open universe at the $3\sigma$
level despite the visual impression given by $\xi_2$, which is due to
the logarithmic axes.  Standard cold dark matter is inconsistent with
$\xi_1$ at many sigma. This first measurement of ellipticity
correlations due to cosmic shear over half-degree angular scales is in
agreement with a variety of other evidence in ruling out standard cold
dark matter.  Weak lens observations of larger fields and more distant
galaxies will be able to clearly distinguish between the remaining
models, or suggest the need for a new model.  }
\end{figure}
\clearpage


\end{document}